# Optically controlled ultrafast dynamics of skyrmion in antiferromagnets

S. H. Guan[1], Y. Liu[1], Z. P. Hou[1], D. Y. Chen[1], Z. Fan[1], M. Zeng[1], X. B. Lu[1], X. S. Gao[1], M. H. Qin[1,*], and J. –M. Liu[1,2]

[1]*Guangdong Provincial Key Laboratory of Quantum Engineering and Quantum Materials and Institute for Advanced Materials, South China Academy of Advanced Optoelectronics, South China Normal University, Guangzhou 510006, China*

[2]*Laboratory of Solid State Microstructures, Nanjing University, Nanjing 210093, China*

**[Abstract]** Optical vortex, an optical beam carrying orbital angular momentum (OAM) has been demonstrated to be promising in manipulating magnetic structure and its dynamics in various systems. In this work, by using numerical and analytical methods, we investigate the optical vortex control of the ultrafast dynamics of skyrmion in antiferromagnets. It is shown that isolated skyrmion can be generated/erased in an ultrashort time of ~ps by beam focusing through the Zeeman effect. Subsequently, the OAM can be transferred to the skyrmion and results in its rotation. Different from the case of ferromagnets, the skyrmion rotation direction in antiferromagnets depends on the light frequency, allowing an easy control of the skyrmion rotation via tuning the frequency of light beam. Furthermore, the skyrmion Hall motion, driven by multipolar spin waves that are excited by optical vortex, is revealed in our calculations, demonstrating the dependence of the Hall angle on the OAM quantum number. This work unveils the interesting optical-control of the skyrmion dynamics in antiferromagnets, which is a crucial step towards the development of magneto-optic and spintronic applications.

---

Email: qinmh@scnu.edu.cn

# I. Introduction

Typically, photons can possess both linear momentum along the propagation direction and spin angular momentum (SAM) related to the circular polarization or chirality property [1-3]. Recently, as a new type of light beam carrying orbit angular momentum (OAM), optical vortex was predicted theoretically [4-8] and realized experimentally using optical elements such as computer-generated holograms, mode conversions, and spiral phase plates [9-11]. Optical vortex has a helical phase wave front which is characterized by an azimuthal phase factor $\exp(im\phi)$ with the OAM quantum number $m$. Moreover, the beam forms a ring-shaped spatial profile of intensity in the cross section because of the zero field topological singularity in vortex core [12,13]. Certainly, such topological feature could be utilized for generating some novel effects in condensed matters.

Indeed, due to its interesting physical characters, optical vortex and its potential applications have received extensive attention. For instance, it has been suggested that optical vortex can be used in super-resolution microscopy [14] and chiral laser ablation [15,16]. Moreover, it may play an important role in modulating particles dynamics via the OAM transfer. Specifically, the OAM transfer from optical vortex to particles drives the rotation of the latter around the beam axis, because the rotating energy flux induced by the Poynting vector propagation exerts a torque onto the particles [17-19]. Thus, optical vortex could be used in optical micromachines like optically driven cogs [19]. Interestingly, besides classical particles, quasiparticles such as magnetic skyrmion [20-25] can also be controlled by optical vortex through the OAM transfer, which is attractive in spintronic applications.

Magnetic skyrmion can be viewed as a topological soliton with noncollinear structure, which may be stabilized by the Dzyaloshinskii-Moriya interaction (DMI) in non-centrosymmetric crystals or by frustrated exchange interactions [26-31]. Owing to the specific characters such as nanoscale in size, nontrivial topology, and low-threshold driving current [32,33], skyrmion is considered to be a promising candidate for information carrier in future spintronic devices. Thus, ultrafast manipulation of skyrmion is one of the most important topics not only in spintronics. Luckily, optical vortex has been revealed to be powerful in modulating effectively magnetic skyrmion. For example, the optical vortex may be coupled with ferromagnets through the Zeeman effect, inducing twisted magnons which in turn favor the generation of skyrmions [16,34]. Furthermore, a vortex beam in interaction with a skyrmion can transfer its OAM and thus drive the skyrmion to rotate around the beam axis [35].

Subsequently, antiferromagnetic (AFM) skyrmions draw attention [36-38], considering that they are free of several disadvantages of ferromagnets including the strong stray field and relatively slow spin dynamics. Unlike skyrmion in ferromagnets, an AFM skyrmion is comprised of two coupled spin configurations with opposite topological numbers, resulting in strong anti-interference capability [39,40]. Besides, it exhibits more desirable properties, such as terahertz oscillation and ultrafast dynamics [41-43].

Undoubtedly, manipulation of an AFM skyrmion using optical vortex becomes an attractive topic which deserves to be explored. First, ultrafast generation of an AFM skyrmion could be realized by applying optical vortex, noting that the AFM dynamics is generally much faster than ferromagnetic one. Second, interesting phenomena induced by the coupling of optical vortex with AFM skyrmion are expected, considering the strong AFM exchange. For example, an additional time inversion symmetry term will be introduced in the spin dynamic equation for antiferromagnets [44], resulting in the dynamic behaviors rather different from those in ferromagnets. At last, optical vortex can excite multipolar spin waves [34], whose interaction with AFM skyrmion in turn influences the skyrmion dynamics. However, the multipolar spin-wave-driven dynamics of AFM skyrmion remains ambiguous, although effective control of AFM skyrmion by plane spin-wave has been uncovered [45], while the underlying mechanisms for the two types of spin wave driven excitation are different.

In this work, we intend to investigate how to manipulate an AFM skyrmion using an optical vortex by performing comprehensive numerical simulation and analytical approach. It will be demonstrated that an isolated skyrmion can be generated/erased in a time as short as ~ps by the optical vortex in the focused beam geometry. Subsequently, the OAM transfer from the vortex to the skyrmion allows the skyrmion rotation and the motion direction depends on the light frequency in addition to the OAM quantum number. This property suggests a convenient scheme in which the skyrmion dynamics is tuned by the light frequency. Furthermore, our simulation suggests that the optical vortex induced multipolar spin waves may generate the skyrmion Hall motion and the Hall angle is dependent of both the light handedness and OAM quantum number. Such an effective control of skyrmion Hall effect is meaningful for spintronic devices, including the skyrmion-based logic units and arithmetic units [46-48], noting that such effect generally hampers the data transportation of devices.

## II. Model and method

Without losing the generality, we start from a two-dimensional (2D) classical Heisenberg model in the $xy$ lattice plane, described by the following Hamiltonian,

$$H = J\sum_{<i,j>}\mathbf{m}_i \cdot \mathbf{m}_j + \sum_{<i,j>}\mathbf{D}\cdot(\mathbf{m}_i \times \mathbf{m}_j) - K\sum_i (m_i^z)^2 - \mathbf{B}\cdot\mathbf{m}_i, \qquad (1)$$

where $\mathbf{m}_i = -\mathbf{S}_i/\hbar$ is the local magnetic moment at site $i$ with local spin $\mathbf{S}_i$ and reduced Planck constant $\hbar$. The first term is the AFM exchange energy between the nearest neighbors with the AFM exchange coefficient $J > 0$, the second term is the bulk DMI with the vector $\mathbf{D} = D\mathbf{e}_{ij}$, $\mathbf{e}_{ij}$ is the unit vector connecting the nearest neighbors, the third term is the anisotropy energy along the $z$-axis with anisotropy coefficient $K$, and the last term is the Zeeman coupling of local magnetic moments with external field $\mathbf{B}$.

We consider a canted antiferromagnetic compound $KMnF_3$ for simulation [38,45,49], which has been widely used in the study of skyrmion dynamics. Therefore, we take the lattice constant $a = 0.5$ nm, the magnetic layer thickness $d = 2$ nm, the coupling constants $J = 1.0 \times 10^{-21}$ J, $D/J = 0.073$, and $K/J = 0.01$ for the subsequent computation. Moreover, due to the non-ferroelectricity character of $KMnF_3$, the effect of electric field can be ignored safely in the Hamiltonian.

The magnetization dynamics is described by solving the Landau-Lifshitz-Gilbert (LLG) equation,

$$\frac{d\mathbf{m}_i}{dt} = -\gamma \mathbf{m}_i \times \mathbf{H}_i^{eff} + \alpha \mathbf{m}_i \times \frac{d\mathbf{m}_i}{dt}, \qquad (2)$$

where $\mathbf{H}_{eff\ i} = -(1/\mu_0)\partial H/\partial \mathbf{m}_i$ is the effective field, $\alpha = 0.01$ is the Gilbert damping coefficient, and $\gamma = -2.211 \times 10^5$ m/(A·s) is the gyromagnetic ratio [38].

Following earlier works, a vortex beam with the Laguerre-Gaussian mode is considered, which carries the following magnetic field profile on the focal plane [34,35],

$$\mathbf{B}(\rho,\phi,t) = B_0(\rho/W)^{|m|}W^{-1/2}e^{-\frac{\rho^2}{W^2}-(\frac{t-t_0}{\sigma})^2+i(m\phi-\omega t)}L_p^{|m|}(2\rho^2/W^2)\mathbf{e}_p, \qquad (3)$$

where $B_0$ is the strength of magnetic field, $W$ represents the beam waist, $\rho$ is the distance between the reference point and the vortex core, $t$ is the relaxation time, $t_0$ determines the peak position of beam intensity, and $\sigma$, $\phi$, $\omega$, $p$ are the beam duration, polar angle, light frequency and the radial index of light, respectively, $\mathbf{e}_p = \mathbf{x} +/- i\mathbf{y}$ corresponds to the left-/right-handed circularly polarized light. For the case of $p = 0$, the generalized Laguerre function $L_p^{|m|}(2\rho^2/W^2) = 1$.

Fig. 1 presents a schematic of the coupling mechanism between the optical vortex and the spin lattice under investigation, where the orange spiral is the light beam with $m = 2$ whose SAM leads to the local magnetization procession, and the OAM induces twisted magnons. Unless stated elsewhere, we set $B_0 = 0.25$ with reduced unit which corresponds to 1.1 Tesla in reality, $W = 5a$ with a spot size ~ 10 nm, $m = -8$, and $\omega = 4$ THz for the optical vortex. It is worth noting that the beam in subwavelength scale can be realized, thanks to the recent development of plasmonics [50].

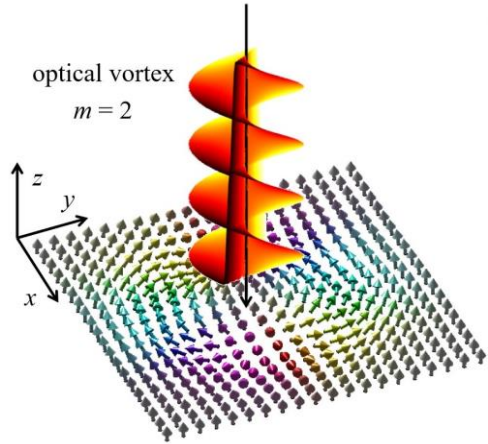

Fig. 1. The coupling mechanism between an optical vortex and a spin lattice, where the orange spiral is the light beam with $m = 2$.

## III. Results and discussion
### A. Ultrafast generation/erasure of isolated AFM skyrmion

First, we investigate the stability of a skyrmion upon interaction with a pulsed optical vortex. Fig. 2(a) shows the evolution of spin configuration for $m = -8$ with a pulse time $\sigma =$ 1.5 ps, which demonstrates the effective skyrmion writing by the optical vortex. Here, $n_z$ is the $z$ component of unit staggered magnetization $\mathbf{n} = (\mathbf{m}_1 - \mathbf{m}_2)/|\mathbf{m}_1 - \mathbf{m}_2|$ with the AFM sublattice magnetizations $\mathbf{m}_1$ and $\mathbf{m}_2$. The vortex induces twisted magnons in AFM system at $t = 2$ ps, and the magnons are coupled due to the DMI, forming a ring-shaped structure ($t = 4$ ps). Subsequently, excessive energy makes the structure evolve into an unstable skyrmionium at $t = 6$ ps, which degenerates into a stable skyrmion finally ($t = 20$ ps). Importantly, the writing process here is independent of device geometry and much faster than those with traditional methods. As a comparison, the skyrmion writing using current pulse [51] and local heating [52] are usually in nanosecond time range. Furthermore, the skyrmion can be easily erased by reversing the sign of $m$ [35,53] through modulating the coupling of chirality and the OAM, as shown in Fig. 2(b), further demonstrating the great potential of optical vortex in manipulating AFM skyrmions.

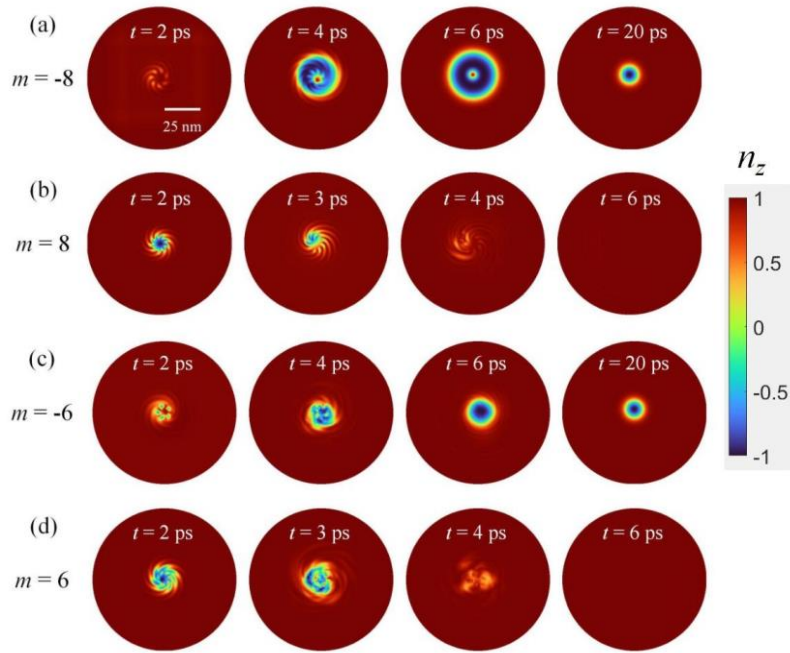

Fig. 2. Ultrafast generation ((a) and (c)) and erasure ((b) and (d)) of isolated AFM skyrmion through optical vortex focusing for ((a) and (b)) $|m| = 8$, $B_0 = 1.1$ T, and ((c) and (d)) $|m| = 6$, $B_0 = 3.1$ T.

As a matter of fact, similar ultrafast writing and erasing processes are observed for other $m$ values, as shown in Figs. 2(c) and 2(d), respectively, where the results for $|m| = 6$ are given. It is noted that a larger $B_0$ is needed to generate/erase the skyrmion for a smaller $|m|$, and the optical vortex with $B_0 = 3.1$ T is used for $|m| = 6$. With the increase of $|m|$, the optical vortex field profile becomes more spatially non-uniform, which leads to the excitation of skyrmion more easily. Furthermore, critical differences may appear for fs, ps, and ns laser pulses. Specifically, ns laser pulse may induce multiple skyrmions, and ps laser pulse induces an isolated skyrmion in the beam focus. For pulse times ~fs, the flipping of magnetic moments is no longer sufficient in such ultra-short response time, and thus no skyrmion can be generated.

**B. AFM skyrmion rotation**

Then, we investigate the OAM transfer from the beam to the skyrmion, which is believed to drive the skyrmion motion. The Lagrangian density can be written as [54],

$$\zeta = \frac{\varepsilon^2}{2A_0}(\dot{\mathbf{n}} - \gamma\mathbf{B}\times\mathbf{n})^2 + D\mathbf{n}\cdot(\nabla\times\mathbf{n}) - \frac{A - L^2/A_0}{2}(\nabla\mathbf{n})^2 + \frac{L\varepsilon}{A_0}\nabla\mathbf{n}\cdot(\mathbf{n}\times\dot{\mathbf{n}}) - Kn_z^2 - \frac{2\gamma\hbar SL}{A_0}\mathbf{B}\cdot\nabla\mathbf{n}, \qquad (4)$$

where $\varepsilon$ is the staggered spin angular momentum density, $A_0$ and $A$ are the homogeneous and inhomogeneous exchange constants, respectively, and $L$ is the parity-breaking constant [54]. For convenience of analytic calculations, one considers the following approximation of the skyrmion configuration [55] in cylindrical coordinates with $\mathbf{n} = (\sin\theta\cos\varphi, \sin\theta\sin\varphi, \cos\theta)$,

$$\begin{aligned}
r &= \sqrt{(\rho\cos\phi - R\cos\phi_0)^2 + (\rho\sin\phi - R\sin\phi_0)^2}, \\
\theta &= 2\arctan[\frac{\sinh(R_s/w)}{\sinh(r/w)}], \\
\varphi &= \frac{\pi}{2} + \frac{\rho\sin\phi - R\sin\phi_0}{|\rho\sin\phi - R\sin\phi_0|}\arccos(\frac{\rho\cos\phi - R\cos\phi_0}{r}),
\end{aligned} \qquad (5)$$

where $R$ is the skyrmion orbit radius, $\phi_0$ is the polar angle of the skyrmion center, $w = \pi D/4K$ is the domain wall width, and $R_s = \pi D[A/(16AK^2 - \pi^2 D^2 K)]^{1/2}$ is the skyrmion radius. Then, the equation describing the motion of skyrmion is obtained,

$$M(\ddot{\mathbf{q}} + \frac{A_0\alpha}{\varepsilon}\dot{\mathbf{q}}) = \mathbf{F}, \qquad (6)$$

where $M$ is the skyrmion effective mass, and $\mathbf{q}$ is the skyrmion coordinate [54]. The tangential force $\mathbf{F}$ contains two terms,

$$F_\omega = \frac{\varepsilon^2 \gamma B_0 \omega}{A_0 \sqrt{W}} \int \frac{(\rho/W)^{|m|} e^{-\rho^2/W^2}}{\rho} [\sin(m\phi \mp \omega t)(\cos\varphi \frac{\partial \theta}{\partial \phi} - \sin\theta \cos\theta \sin\varphi \frac{\partial \varphi}{\partial \phi})$$
$$\pm \cos(m\phi \mp \omega t)(\sin\varphi \frac{\partial \theta}{\partial \phi} + \sin\theta \cos\theta \cos\varphi \frac{\partial \varphi}{\partial \phi})]dV$$

$$F_m = \frac{\varepsilon \gamma L B_0 m}{A_0 \sqrt{W}} \int \frac{(\rho/W)^{|m|} e^{-\rho^2/W^2}}{\rho^2} [-\sin(m\phi \mp \omega t + \Phi)(\cos\theta \sin\varphi \frac{\partial \theta}{\partial \phi} + \sin\theta \cos\varphi \frac{\partial \varphi}{\partial \phi})$$
$$\pm \cos(m\phi \mp \omega t + \Phi)(\cos\theta \cos\varphi \frac{\partial \theta}{\partial \phi} - \sin\theta \sin\varphi \frac{\partial \varphi}{\partial \phi})]dV$$

(7)

where the sign + (−) in ± corresponds to the left- (right-) handed light. Here, $F_\omega$ and $F_m$ come from the temporal and spatial deflection of the beam profile, respectively. A phase $\Phi$ term is introduced into $F_m$ to ensure that the two forces are always asynchronous.

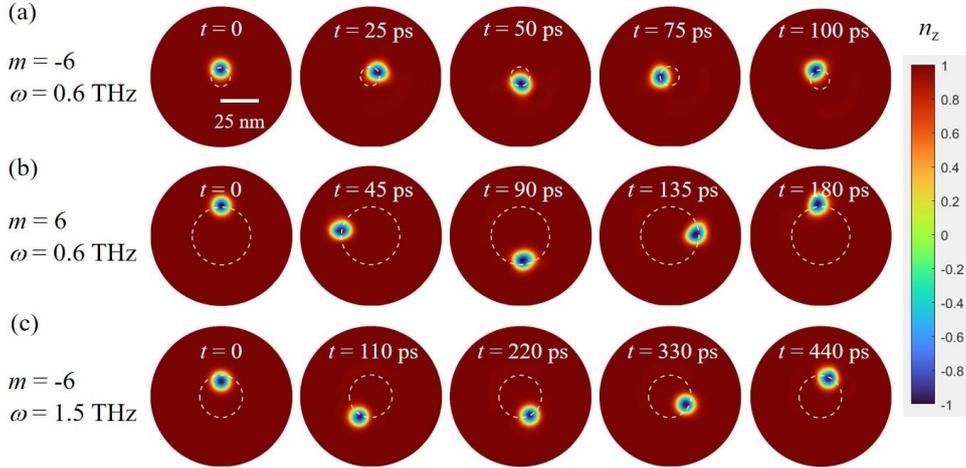

Fig. 3. Skyrmion gyration driven by the optical vortex with (a) $m = -6$, and $\omega = 0.6$ THz, (b) $m = 6$, and $\omega = 0.6$ THz, and (c) $m = -6$, and $\omega = 1.5$ THz. The white dotted circles are the trajectories of the skyrmion, whose radius $R$ depends on both $m$ and $\omega$.

In Fig. 3(a), we present the LLG-simulated trajectory of the AFM skyrmion driven by the left-handed continuous-wave with $m = -6$, $B_0 = 1.1$ T, $\omega = 0.6$ THz, and $W = 25a$ equal to a spot size ~ 50 nm. Like skyrmion in ferromagnets, the AFM skyrmion is driven by the optical vortex rotating around the core with a very high angular frequency under the restraint of the optical potential well [35]. Here, the skyrmion rotates clockwise in the oscillation and breathing modes. Furthermore, a reversed $m$ generates an opposite OAM of the light, resulting in an anti-clockwise rotation of the skyrmion, as shown in Fig. 3(b), where the skyrmion trajectory for *m* = 6 is presented. Besides the rotation direction, both the orbit radius and speed are significantly changed. Interestingly, the rotation direction can also be controlled by modulating frequency $\omega$. For example, for *m* = −6, an anti-clockwise rotation is realized when $\omega$ increases beyond 1.5 THz, as shown in Fig. 3(c).

To further explore this attractive phenomenon, the average tangential velocity *v* of the skyrmion as a function of $\omega$ for various *m* are summarized in Fig. 4(a). There is a frequency threshold $\omega$ ~ 0.4 THz beyond which the skyrmion can be effectively driven. Then, the rotation speed first increases and then decreases with increasing frequency, and it reaches a maximum value ~900 m/s around $\omega$ = 0.5 THz, which is much faster than that in the ferromagnetic skyrmion [35]. Importantly, there exists a critical frequency $\omega_c$ ~ 1.0 THz, above which the rotation direction is changed from clockwise to anticlockwise due to the reverse of the OAM, which is different from the case of ferromagnets where the rotation direction can hardly be affected by frequency.

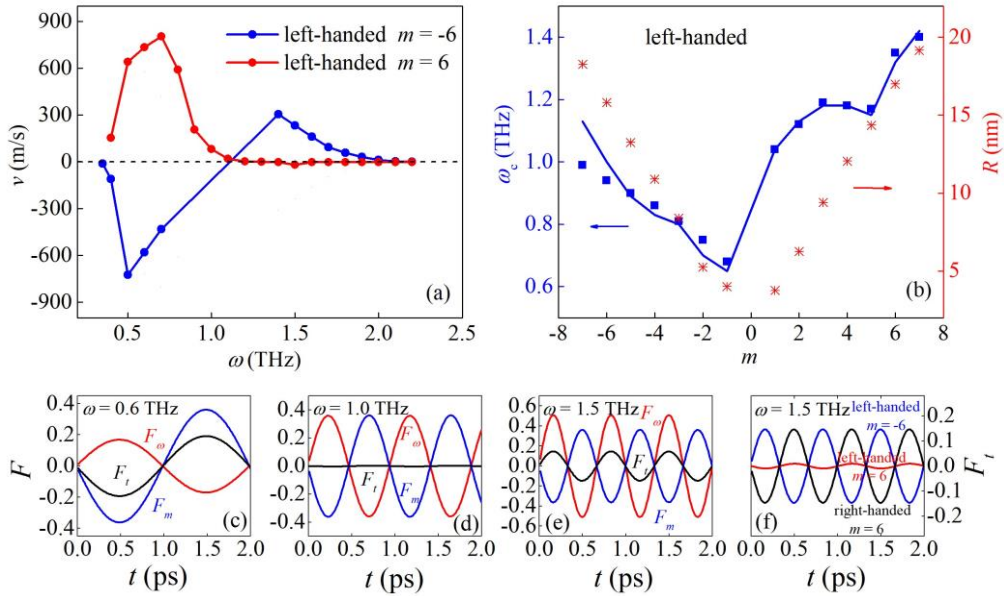

Fig. 4. (a) The skyrmion tangential velocity *v* as a function of $\omega$ for opposite values of *m*, and (b) the numerically simulated (blue solid squares) and analytically calculated (blue line) critical frequency $\omega_c$ for various *m*. The skyrmion orbit radiu *R* is also presented by red star. (c)-(e) Analytically calculated tangential forces $F_\omega$, $F_m$ and $F_t$ for various $\omega$, and (f) $F_t$ for different light handedness and *m* at $\omega$ = 1.5 THz.

In some extent, the reversal of skyrmion rotation with increasing $\omega$ can be understood qualitatively from the competition between $F_\omega$ and $F_m$, which come from respectively the time-varying magnetic field and spatial non-uniform magnetic field. For $\omega = 0.6$ THz below $\omega_c$, the magnitude of $F_m$ is larger than that of $F_\omega$, as shown in Fig. 4(c) where the calculated $F_\omega$, $F_m$, and $F_t = F_\omega + F_m$ as functions of time are presented, resulting in the clockwise rotation of the skyrmion. The magnitude of $F_\omega$ increases with the increase of $\omega$, while $F_m$ can hardly be affected by $\omega$. Thus, when the two forces are cancelled out each other at $\omega \sim 1.0$ THz, as shown in Fig. 4(d), the skyrmion rotation is completely suppressed. For $\omega > \omega_c$, $F_\omega$ overwhelms $F_m$, as shown in Fig. 4(e), resulting in an opposite OAM and counter-rotation of the skyrmion. A further increase of $\omega$ from 1.5 THz speeds down the skyrmion gradually to zero, due to the mismatch between the optical frequency and intrinsic frequency of the AFM skyrmion.

The critical frequency $\omega_c$ depends on both the quantum number $m$ and orbit radius $R$, which can be analytically estimated from Eq. (7). The calculated and simulated $\omega_c$ as functions of $m$ are given in Fig. 4(b), and they coincide well with each other. A larger orbit radius is expected for a larger $|m|$, due to the fact that the peak position of the beam intensity shifts far away from the beam center as $|m|$ increases, as confirmed in the simulations. Above $\omega_c$, the rotation speed of the skyrmion depends on the sign of $m$. For example, the speed of the skyrmion for $m = 6$ is one order of magnitude lower than that for $m = -6$, as shown in Fig. 4(a). This phenomenon attributes to the fact that the OAM transfer from the beam to the skyrmion depends on the coupling of the light OAM and local magnetization precession which is related to the beam chirality. In the case of the left-handed light with $m = -6$, the light OAM and magnetization precession are with the same direction, resulting in a strong OAM transfer. Conversely, the OAM transfer is extensively suppressed for $m = 6$ due to the opposite directions of the light OAM and the magnetization precession. Thus, both the magnitude of $F_t$ and the rotation speed of the skyrmion for $m = -6$ are much larger than those for $m = 6$, as shown in Fig. 4(f).

The above analysis can be easily applied to the discussion on the skyrmion dynamics driven by right-handed light, based on symmetry analysis. Specifically, the right-handed light with *m* generates a force $F_t$, opposite to that of the left-handed light with $-m$, as shown in Fig. 4(f) where $F_t$ induced by the right-handed light with $m = 6$ is also presented. As a result, the skyrmions in the two cases rotate reversely with the same rotation speed. On the other hand, similar driving effects for linearly polarized lights with $+m$ and $-m$ are revealed due to the absence of SAM (not shown here), further confirming the above analysis.

By solving Eq. (6), the skyrmion velocity is evaluated to be,

$$v = \frac{C_1 \varepsilon B_0 F_A}{M A_0 \alpha W (A_0^2 \alpha^2 / \varepsilon^2 + \omega^2)} \omega^{C_2} e^{-C_3(\omega - \omega_c)}, \tag{8}$$

where $F_A$ is the amplitude of $F_t$, $C_1 = 4.8$, $C_2 = 2$, and $C_3 = 2$ are the fitting coefficients estimated from the LLG-simulated results in Fig. 4(a).

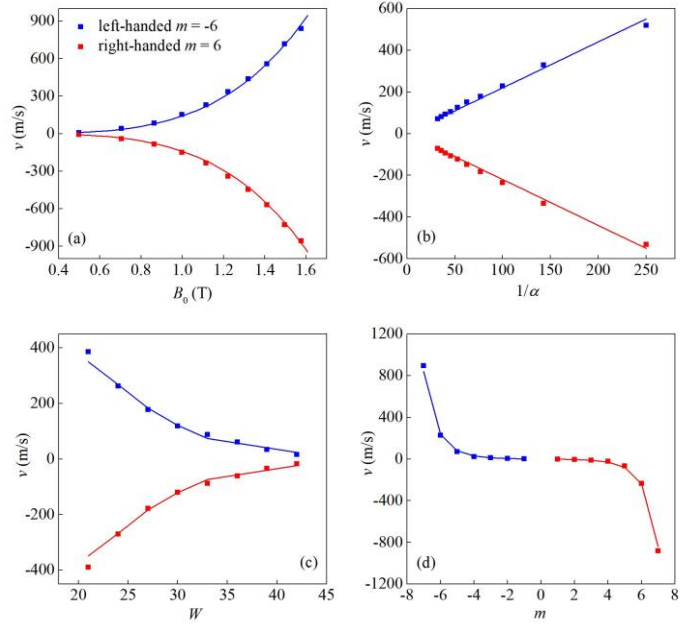

Fig. 5. The simulated (solid squares) and analytically fitted (sold lines) *v* as functions of (a) beam intensity $B_0$, (b) damping coefficient $\alpha$, (c) beam waist *W* for left-handed light with $m = -6$ and right-handed light with $m = 6$, and (d) OAM quantum number *m*.

Furthermore, the dependences of the skyrmion velocity on several physical parameters including $\alpha$, $B_0$, *W* and *m* are investigated. By Eq. (8), the calculated (solid lines) and LLG-simulated (solid symbols) velocities as functions of field $B_0$, damping constant $\alpha$, and beam waist *W*, driven by the left-handed beam with $m = -6$ and right-handed beam with $m = 6$ are plotted in Figs. 5(a)-5(c), demonstrating the good coincidence between the theoretical analysis and simulations. First, $v \propto B_0^2$ is obtained because the magnetic energy density linearly depends on $B_0^2$. Second, an enhanced damping term always reduces the skyrmion mobility, and the velocity is inversely proportional to $\alpha$. Furthermore, as beam waist $W$ increases, the rotation radius $R$ is increased, while the beam energy density is decreased, which speeds down the skyrmion. Fig. 5(d) shows the skyrmion velocity as a function of $m$ and demonstrates the decrease of $v$ with decreasing $|m|$, which is well consistent with Eq. (8).

**C. AFM skyrmion Hall motion driven by light vortex-induced multipolar spin waves**

It is noted that vortex beam can also excite multipolar spin waves in antiferromagnets, and their interaction with the AFM skyrmion remains ambiguous. Thus, for the integrity of this work, we also investigated the interaction between the AFM skyrmion and multipolar spin waves. Here, we consider a continuous light beam with high frequency and reset the parameters to be $W = 5a$ and $\omega = 3.5$ THz to excite spin waves. The beam focuses on a position far away from the skyrmion to avoid the directly influence of the beam profile, as depicted in Fig. 6(a).

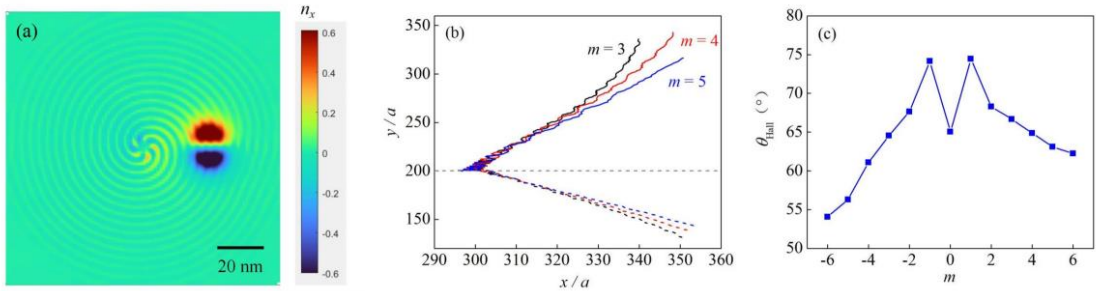

Fig. 6. (a) The interaction between the skyrmion and multipolar spin waves induced by the optical vortex with $m = 5$, where the light-colored cylindrical wave profiles refer to the spin waves while the dark-colored profile refers to the skyrmion. (b) The skyrmion trajectories driven by multipolar right-handed (solid lines) and left-handed (dashed lines) spin waves with various $m$, and (c) the dependence of the skyrmion Hall angle $\theta_{\text{Hall}} = \arctan(v_y/v_x)$ on $m$. The horizontal gray dashed line in (b) represents the wave vector direction.

The spin wave mode has the following cylindrical wave form,

$$\Psi = \frac{1}{\sqrt{\rho}} \exp[i(k\rho + m\phi - \omega t)], \quad (9)$$

where $k$ is the wave number whose direction depends on angle $\phi$. Then, the scattering amplitude of the spin waves is given by [56],

$$f(\chi) = \frac{e^{-i\pi/4}}{\sqrt{2\pi k}} \sum_{\substack{l=-\infty \\ l \neq -m}}^{\infty} \frac{e^{-il\pi/2}}{m+l}(1-e^{i2\delta_l})e^{-i[(m+l)\chi+l\pi/2]}, \tag{10}$$

where $\delta_l$ is the phase shift with the partial waves index $l$, and $\chi$ is the scattering angle related to the wave vector.

It is noted that the OAM quantum number $m$ couples with $l$ and the phase shift, thus it may affect the scattering amplitude distribution and break the degeneracy between the left-handed and right-handed magnons (the quantum of spin wave). The assumption is verified by the simulations, as shown in Fig. 6 (b), where the skyrmion trajectories driven by the multipolar right-handed (solid lines) and left-handed (dashed lines) spin waves with various $m$ are presented. An obvious asymmetry of skyrmion trajectories for right- and left-handed spin waves is observed, which is different from the case of the plane spin waves [45,57].

It is shown that the multipolar spin waves can drive the skyrmion Hall motion deviating from the wave vector direction (along the $x$-axis), similar to the case of plane spin waves. Since magnons may be regarded as charged particles, their signs are opposite for the left-handed and right-handed spin waves. The particles suffer a fictitious magnetic field induced by the skyrmion topology, leading to the fact that the spin waves with opposite handedness are scattered by the skyrmion toward opposite directions [57]. As a result, the skyrmion can be effectively driven by the spin waves through the momentum exchange, and the transverse direction depends on the wave handedness.

Interestingly, the Hall motion of the skyrmion can be affected by the OAM quantum number $m$. In Fig. 6(c) is plotted the dependence of skyrmion Hall angle $\theta_{\text{Hall}} = \arctan(v_y/v_x)$ on $m$, which demonstrates the suppression of the Hall motion with the increase of $|m|$ attributing to the increase of magnons scattering angle. Thus, one may tune the OAM quantum number to modulate the skyrmion Hall motion, which is very meaningful for skyrmion-based spintronic applications.

D. **Discussion**

To this end, it is clearly shown that the control of skyrmion by the optical vortex is efficient and easy-accessed. Given the fact that nanoscale control of magnetic textures is an important issue with great challenges, the efficient control of magnetic system using plasmonic materials and vortexes, is highly appreciated, as also revealed in earlier works [58-60].

In this work, it has been revealed that the AFM skyrmion can be manipulated in tempo-spatial nanoscales via the focusing optical vortex. Comparing to traditional schemes such as spin polarized current driven method, magnetic texture control using vortex beam has many advantages including more adjustable degrees of freedom, shorter pulse time, and lower dissipation. The ultrafast generation/erasure and dynamics of the AFM skyrmion induced by optical vortex are demonstrated here, using numerical and analytical methods. The present wok is certainly an important step toward updated spintronic device design based on AFM skyrmion and may contribute to the development of magneto-optical memories and logic devices with high integration and speed. Furthermore, the optically controlled ultrafast dynamics of other types of spin defects, such as magnetic vortex [61] and bimeron [62] in frustrated magnets and ferrimagnets could also be expected.

It is noted that, a THz light beam with wavelength ~ $\mu$m is used in our simulation, and a nanoscale beam waist $W$ cannot be realized in the free space because of the diffraction limit. However, the main results and conclusion is independent of the system size, qualitatively, at least, which can be safely transferred to large systems. Specifically, in Section IIIA, a $W$ comparable to the skyrmion size ~ $\mu$m can generate/erase skyrmions through magnetization switch. In Section IIIB, In Section IIIB, different from the light frequency $\omega$ and OAM quantum number $m$, the beam waists $W$ hardly affects the OAM transfer mechanism as demonstrated in Eq. (7), and there is not a strict requirement for the selection of $W$. The spot size increases and the beam energy decreases with the increasing $W$, and a stronger OAM of light is needed to maintain the skyrmion speed. In Section IIIC, the excitation of multipolar spin waves is rather easy and independent of $W$, because they are long-wavelength phenomena [34].

Moreover, THz optical vortex can be generated via various techniques such as spiral phase plates [10,63], and nano-focusing of the beam can be realized with the help of metal antenna tips or plasmonic waveguides formed by dielectric nanowires [49,64], ensuring that the parameters of the light beam are physically achievable. The Laguerre-Gaussian mode i.e. Eq. (3) forms a complete and orthogonal basis set [65,66] based on which the focused light beam can be also described. Besides, subwavelength focusing of optical vortex by metallic nanoscale resonant optical antennas has been revealed both theoretically and experimentally [9,63]. Importantly, it is demonstrated in the Lumerical simulations that an antenna consisting of a larger number of arms and having higher symmetry would be able to reproduce the field profile described by Eq. (3) more stably [9]. Thus, the prediction in this work is reliable and does provide critical information for optical control AFM skyrmions, which deserves to be checked in future experiments.

## IV. Conclusion

In conclusion, we have studied numerically and analytically the skyrmion dynamics in antiferromagnets driven by the optical vortex. It has been demonstrated that the vortex beam can be employed to excite the twisted magnons and generate/erase the skyrmion in a time-scale as short as ~ps. The OAM transfer from the light to the skyrmion results in the rotation of the skyrmion, whose direction can be modulated by tuning the light frequency in addition to the OAM quantum number. This interesting behavior allows one to modulate the skyrmion rotation easily through tuning the frequency. Furthermore, the optical vortex can excite multipolar spin waves, which in turn drive the skyrmion Hall motion. The Hall angle also depends on the OAM quantum number, providing a new degree of freedom to better control the skyrmion motion.

**Acknowledgment**

This work is supported by the Natural Science Foundation of China (Grants No. U22A20117, No. 51971096, No. 92163210, and No. 51721001), the Guangdong Basic and Applied Basic Research Foundation (Grant No. 2022A1515011727), and Funding by Science and Technology Projects in Guangzhou (Grant No. 202201000008).

.